\RequirePackage{fix-cm}
\documentclass[tex,tightenlines,twocolumn,epjc3,tightenlines]{svjour3}
\smartqed  
\usepackage{comment}
\usepackage{graphicx}
\usepackage{graphics}
\usepackage{amsmath, amssymb}
\usepackage{graphics,bm}
\usepackage{graphicx}
\usepackage{bbold}
\usepackage{slashed}
\usepackage{feynmf}
\usepackage{wrapfig}
\usepackage{widetext}
\usepackage{hyperref}
\usepackage{cancel}
\usepackage[usenames]{color}     

\newcommand{\be}{\begin{equation}}
\newcommand{\ee}{\end{equation}}

\newcommand{\bqa}{\begin{eqnarray}}
\newcommand{\eqa}{\end{eqnarray}}

\def\square{\vcenter{\vbox{\hrule height.4pt
          \hbox{\vrule width.4pt height4pt
          \kern4pt\vrule width.3pt}\hrule height.4pt}}}
  
\begin{document}
\title{
Quark, pion and axial condensates in three-flavor finite isospin
chiral perturbation theory}

\author{Prabal Adhikari\thanksref{e1,addr1,addr2} \and Jens O. Andersen\thanksref{e2,addr2}\and Martin A. Mojahed\thanksref{e3,addr2}}
\thankstext{e1}{e-mail: adhika1@stolaf.edu}

\thankstext{e2}{e-mail: andersen@tf.phys.ntnu.no} 
\thankstext{e3}{e-mail: martimoj@stud.ntnu.no} 
\institute{St. Olaf College, Faculty of Natural Sciences and Mathematics, Physics Department, 1520 Saint Olaf Avenue,
  Northfield, MN 55057, United States 
  \label{addr1}
          \and
  Department of Physics, 
Norwegian University of Science and Technology, H{\o}gskoleringen 5,
N-7491 Trondheim, Norway
          \label{addr2}
}            

\maketitle

\begin{abstract}
We calculate the light-quark condensate, the strange-quark condensate, the pion condensate, and the axial condensate in three-flavor chiral perturbation theory ($\chi$PT) in the presence of an isospin chemical potential at next-to-leading order at zero temperature. 
It is shown that the three-flavor $\chi$PT effective potential and condensates can be mapped onto two-flavor $\chi$PT ones by integrating out mesons with strange quark content (kaons and eta), with renormalized couplings. 
We compare the results for the light quark
and pion condensates at finite pseudoscalar source with ($2+1$)-flavor lattice QCD, and we also compare the axial condensate at zero pseudoscalar and axial sources with lattice QCD data. 
We find that the light quark, pion, and axial condensates are in very good agreement with lattice data. There is an overall improvement by including NLO effects.

\end{abstract}

\section{Introduction}
\label{intro}
Quantum Chromodynamics (QCD) has a rich phase structure, which can be established using its symmetries and symmetry breaking patterns~\cite{raja,alford,fukurev}. The QCD Lagrangian possesses an $SU(N_{c})$ gauge symmetry with $N_{c}=3$, which preserves color charge when quarks and gluons interact. Furthermore, the QCD Lagrangian is symmetric with respect to independent chiral rotations of left-and-right handed quarks and anti-quarks in the chiral limit.  However, the QCD vacuum breaks this symmetry by pairing quarks and antiquarks giving rise to a non-zero, spatially homogeneous, chiral condensate, $\langle\bar{\psi}\psi\rangle_{0}$, with the following spontaneous symmetry breaking pattern~\cite{Nambu}
\begin{equation}
\begin{split}
SU(3)_L\times SU(3)_R\times U(1)_B\rightarrow SU(3)_V\times U(1)_B\ .
\end{split}
\end{equation}
Here $SU(3)_{L(R)}$ is the symmetry group associated with chiral transformations of left(right)-handed quarks in the chiral limit and $SU(3)_{V}$ is the symmetry group associated with vector transformations of the quarks.~\footnote{For a thorough review, see for instance \cite{Scherer} and \cite{mannarev}.} The pairing is analogous to Cooper pairs in the Bardeen-Cooper-Schrieffer (BCS) theory of superconductivity~\cite{BCS} as was originally pointed out by Nambu~\cite{Nambu}. $U(1)_{B}$ is the symmetry of the QCD Lagrangian with respect to global phase transformations of quarks and anti-quarks, which leads to the conservation of baryon charge. The order parameter of spontaneous chiral symmetry breaking is the chiral condensate, $\langle\bar{\psi}\psi\rangle_{0}$, which is non-zero in the QCD vacuum and leads to the meson octet of (pseudo-) Nambu-Goldstone bosons for (massive) massless quarks. Since quarks are massive, with the strange quark mass being much larger than the up and down quark masses, the vector symmetry group $SU(3)_V$ is broken down to $SU(2)_I\times U(1)_{Y}$ in the isospin limit with $m_{s}\gg m_{u}=m_{d}$.

The presence of isospin and strange chemical potentials in the vacuum phase explicitly breaks the symmetry down as follows,
\begin{align}
SU(2)_I\times U&(1)_Y\times U(1)_B\\
\nonumber
\tfrac{\mu_{I}\neq 0}{}&\downarrow\tfrac{\mu_{s}\neq 0}{}\;,\\
 U(1)_{I_3}\times U&(1)_Y\times U(1)_B\; ,
\end{align}
where $Y$ represents hypercharge, $I$ represents isospin, and $I_{3}$ the third component of isospin. The electromagnetic gauge group $U(1)_{Q}$ is a subgroup of both $SU(3)_L\times SU(3)_R$ and $SU(2)_I\times U(1)_Y\times U(1)_B$ as would be expected from the fact that quarks carry electromagnetic charge in addition to color charge.  
In this work, we will maintain a focus on the properties of QCD at low energies for finite isospin chemical potential~\cite{son}. For  isospin chemical potentials larger than the pion mass (at zero temperature), QCD is known to exhibit pion condensation. This is signaled by the formation of a pseudoscalar condensate $\langle \bar{\psi}\gamma_{5}\tfrac{\lambda_{1,2}}{2}\psi\rangle\neq 0$, where $\lambda_i$ denotes the $i$'th Gell-Mann matrix. This condensate becomes inhomogeneous in the presence of an external magnetic field due to the spontaneous symmetry breaking of the $U(1)$ gauge symmetry~\cite{isomag}. Furthermore, the condensate increases monotonically as has been observed in lattice QCD and NLO two-flavor calculations~\cite{2fcondensate} for values of the isospin chemical potential up to approximately $2m_{\pi}$. This behavior is analogous to that observed in the context of nuclear matter~\cite{Cohen} in that there is a simultaneous weakening of the chiral condensate with increasing nuclear density or in our case isospin density.

There is a further condensate: the axial condensate, which is non-zero in the context of finite isospin chemical potential. It has been studied in lattice QCD with the added benefit that unlike the pion and chiral condensates, the zero-source limit results have been extracted~\cite{pionstar}. We have previously compared finite source results for the pion and chiral condensates with two-flavor QCD~\cite{2fcondensate} with results in very good agreement.

The axial condensate condenses simultaneously with the pion condensate but also exhibits the feature that it does not increase monotonically with increasing isospin chemical potential even though the pion condensate does. Tree-level $\chi$PT calculations show that the condensate increase steadily at the critical isospin chemical potential and peaks at $3^{1/4}m_{\pi}$~\cite{axialfirst} and for larger isospin chemical potentials decreases monotonically. The condensation of the axial condensate is somewhat surprising in that it occurs even in the absence of an explicit axial chemical potential as has also been shown in the NJL model~\cite{NJLaxial}. A somewhat heuristic argument for the condensation was first put forth in Ref.~\cite{axialfirst} using the soft-pion theorem valid for any local operator $\hat{O}$,

\begin{equation}
\label{softpion}
    \lim_{p\rightarrow 0}\langle\pi_{a}(p)s_{1}|\hat{O}|s_{2}\rangle=\frac{i}{\tilde{f}_{\pi}}\langle s_{1}|[\hat{O},\hat{Q}_{a}^{5}]|s_{2}\rangle\;,
\end{equation}
which relates a matrix element involving two arbitrary states $|s_{1}\rangle$ and $|s_{2}\rangle$ with one that involves a new state $|\pi^{a}(p)s_{1}\rangle$ containing an extra pion compared to $|s_{1}\rangle$. $\hat{Q}^{5}_{a}$ is the axial charge operator, which can be defined in terms of the axial charge density operator $\hat{\rho}_{a}^{0}(\vec{y})$,
\begin{align}
    \hat{\rho}_{a}^{0}(\vec{y})&=\bar{\psi}(\vec{y})\gamma^{0}\gamma^{5}\tfrac{\lambda_{a}}{2}\psi(\vec{y})\;,\\
\hat{Q}^{5}_{a}&=\int d^{3}y\ \hat{\rho}^{0}_{a}(\vec{y})\;.
\end{align}
Finally, $\tilde{f}_{\pi}$ is the relevant pion decay constant, which depends on the choice of states $|s_{i}\rangle$, which we will choose to be the pion condensed vacuum that forms in the presence of an isospin chemical potential. The rotation from the normal vacuum, $|0\rangle$, to a pion-condensed vacuum, $|\alpha\rangle$, occurs above a critical chemical potential equal to the pion mass, with the parameter $\alpha$ depending on the isospin chemical potential. The chiral condensate is non-vanishing in both of the vacua. Additionally, in the pion-condensed vacuum $|\alpha\rangle$ both the isospin density and the pion condensate are non-vanishing, i.e.
\begin{align}
    \langle \alpha|\hat{n}_{I}(\vec{x})|\alpha\rangle&\neq 0\;,\\
    \langle \alpha|\hat{\pi}_{a}(\vec{x})|\alpha\rangle&\neq 0\;,    
\end{align}
where $\hat{n}_{I}$ is the isospin density operator and pion condensate (operator) are defined as
\begin{align}
    \hat{n}_{I}(\vec{x})&=\bar{\psi}(\vec{x})\gamma^{0}\tfrac{\lambda_{3}}{2}\psi(\vec{x})\;,\\
    \label{question}
    \hat{\pi}_{a}(\vec{x})&=\bar{\psi}(\vec{x})\gamma^{5}\tfrac{\lambda_{a}}{2}\psi(\vec{x})\; .
\end{align}
Using standard equal-time anti-commutation relations
for the quark fields, it is straightforward to show that
\begin{equation}
\label{axialdensity}
    \Big [\hat{n}_{I}(\vec{x}),\hat{Q}^{5}_{\pm} \Big ]=\pm\hat{\rho}^{0}_{\pm}(\vec{x})\;,
\end{equation}
where $\hat{Q}^{5}_{\pm}=\hat{Q}_{1}^{5}\pm i\hat{Q}_{2}^{5}$ is the charge associated with $\hat{\rho}^{0}_{\pm}=\hat{\rho}_{1}^{0}(\vec{x})\pm i\hat{\rho}_{2}^{0}(\vec{x})$ and $\hat{\rho}^{0}_{\pm}$ are the axial current density operators, $\bar{u}\gamma^{0}\gamma^{5}d$ and $  \bar{d}\gamma^{0}\gamma^{5}u$ respectively. Choosing $|s_{1}\rangle=|s_{2}\rangle=|\alpha\rangle$ and $\hat{O}=\hat{n}_{I}$ in the soft-pion theorem Eq.~(\ref{softpion}) and noting that in the thermodynamic limit, adding a single zero-momentum pion to the pion condensed vacuum does not alter it, i.e. $|\pi^{a}(0)\alpha\rangle=|\alpha\rangle$, we get using Eq.~(\ref{axialdensity}) that the axial density of the pion condensed phase is non-zero, i.e. $\langle\alpha|\hat{\rho}^{0}_{\pm}|\alpha\rangle\neq 0$~\cite{axialfirst}.

The paper is organized as follows: In section \ref{Lagrangian}, we discuss the $\chi$PT Lagrangian in the presence of a pseudo-scalar source and an axial vector potential and point out that the pion condensate and the axial condensate condenses orthogonally in the ground state at tree level. In section~\ref{Veffective}, we construct the one-loop effective potential in the presence of both a pseudo-scalar and an axial vector potential using the ingredients of the previous section. Using the effective potential, we calculate the chiral condensate, the strange-quark condensate, the pion condensate, and the axial condensate in section 4. 
We also map our three-flavor $\chi$PT results to two-flavor $\chi$PT with appropriate identifications of the low energy constants (LECs). Finally, in section~\ref{numerics} we compare the condensates, in particular the axial condensate at zero pionic source and the pion and chiral condensates at finite pionic source with the available lattice data. We list a few useful formulas in Appendix A.

\section{$\chi$PT Lagrangian}
\label{Lagrangian}
$\chi$PT is a low-energy effective theory for QCD based on its symmetries and
degrees of freedom~\cite{wein,gasser1,gasser2,bein}. For two-flavor QCD,
the degrees of freedom are the pion triplet, whereas for three-flavor QCD, they
are the octet of pions, kaons, and the eta.
The leading-order term in $\chi$PT is given by
the following Lagrangian 
\bqa
\mathcal{L}_{2}={f^2\over4}{\rm Tr}
  \left [\nabla_{\mu} \Sigma^{\dagger} \nabla^{\mu}\Sigma     \right ]
+{f^2\over4} {\rm Tr}
 \left [ \chi^{\dagger}\Sigma+\Sigma^{\dagger}\chi\right ]\;,
\label{lag0}
\eqa
where $\chi$ is 
\begin{equation}
\begin{split}
\chi={2B_0M}+{2iB_0j_{1}}\lambda_1+{2iB_0j_{2}}\lambda_2
\end{split}
\end{equation}
with $M={\rm diag}(m_u,m_d,m_s)$ being the quark mass matrix, $j_1$ and $j_2$
are pseudo-scalar (pionic) sources and $\lambda_{i}$ are the Gell-Mann matrices. We will work in the isospin limit, i.e. $m_{u}=m_{d}$ in this paper. The Gell-Mann matrices are normalized as
\begin{equation}
\begin{split}
\tfrac{1}{2}{\rm Tr}(\lambda_i\lambda_j)=\delta_{ij}\;.
\end{split}
\end{equation}
Finally, the covariant derivatives contain both a vector source $v_{\mu}$ and an axial source $a_{\mu}$
\bqa
\nabla_{\mu} \Sigma&\equiv&
\partial_{\mu}\Sigma-i\left [v_{\mu},\Sigma \right]-i\{a_{\mu},\Sigma\}\ ,\\ 
\nabla_{\mu} \Sigma^{\dagger}&\equiv&
\partial_{\mu}\Sigma^{\dagger}-i [v_{\mu},\Sigma^{\dagger} ]+i\{a_{\mu},\Sigma^{\dagger}\}\;,
\eqa
with 
\begin{equation}
\begin{split}
v_{\mu}&=v_{0}^{a}\tfrac{\lambda_{a}}{2}\delta_{\mu}^{\ 0},\ v_{0}^{3}=\mu_{I}\;,\\
a_{\mu}&=a_{0}^{a}\tfrac{\lambda_{a}}{2}\delta_{\mu}^{\ 0}\;,
\end{split}
\end{equation}
where $\mu_{I}$ is the isospin chemical potential and $a_{0}^{a}$ is the zeroth-component of the axial source that couples to $\lambda_a$. 

In two-flavor QCD, 
the presence of an isospin chemical potential rotates the vacuum in the $\tau_{1}$ and $\tau_{2}$ directions~\cite{son}, 
\bqa
\Sigma_{\alpha}&=&e^{i\alpha(\hat{\phi}_1\tau_1+\hat{\phi}_2\tau_2)}\;,
\eqa
with $\hat{\phi}_1$ and $\hat{\phi}_2$ being real parameters 
satisfying $\hat{\phi}_1^2+\hat{\phi}_2^2=1$
such that the ground
state is unitary and properly normalized, i.e. $\Sigma^{\dagger}_{\alpha}\Sigma_{\alpha}=\mathbb{1}$. 
In three-flavor QCD, the vacuum is rotated
in the same way~\cite{kogut3,us} but with $\tau_{1}$ and $\tau_{2}$ replaced by $\lambda_{1}$ and $\lambda_{2}$,
\bqa
\Sigma_{\alpha}&=&e^{i\alpha(\hat{\phi}_1
\lambda_1+\hat{\phi}_2\lambda_2)}\;.
\label{notrotten}
\eqa
The pions then condense in the 
$\begin{pmatrix}
\hat{\phi}_1&,\hat{\phi}_2\\
\end{pmatrix}^{T}$-direction in isospin space. As suggested by the heuristic argument in the previous section, the
pion condensate induces an axial condensate that points in the orthogonal direction,
$\begin{pmatrix}
-\hat{\phi}_2&,\hat{\phi}_1\\
\end{pmatrix}^{T}$.
This feature has also been observed in the context of the NJL model~\cite{NJLaxial}
and previously in $\chi$PT near the critical isospin chemical potential~\cite{LoewefiniteT} at next-to-leading order.

As such we proceed, without any loss of generality, by choosing
$\hat{\phi}_1=0$, $\hat{\phi}_2=1$ and $j_1=0$, $j_{2}=j$ in the following discussion.
The static Lagrangian, which is equal to the tree-level effective potential modulo a minus sign, in three-flavor QCD is
\bqa
\mathcal{L}_{2}^{\rm static}&=&\frac{f^{2}}{2}\Big [
4B_{0}(m\cos\alpha+j\sin\alpha)+
2B_{0}m_{s}
\nonumber
\\ &&
+(a_0^1\cos\alpha+\mu_I\sin\alpha)^2\Big]\;.
\eqa
Since the pion and axial condensates are derivatives 
with respect to the sources $j_{i}$ and $a_{0}^{a}$, respectively, we can immediately deduce from the tree-level effective potential that the tree-level pion and axial condensates are orthogonal as expected. However, this is not sufficient to guarantee that orthogonality holds at next-to-leading order. In order to verify this, one needs to construct the full dispersion relation (including the most general pionic and axial sources) that determines that NLO effective potential. While the full dispersion relation is too cumbersome to present here, we have explicitly verified that this is indeed the case.
With this understanding, we proceed by writing down the rotated vacuum in the pion condensed phase assuming all of the pion condensate points in the $\lambda_{2}$ direction and the axial condensate points in the $\lambda_{1}$ direction. Using Eq.~(\ref{notrotten}), we get for the rotated vacuum 
\bqa
\Sigma_{\alpha}
&=&{1+2\cos{\alpha}\over3}\mathbb{1}+i\lambda_2\sin\alpha
+{\cos\alpha-1\over\sqrt{3}}\lambda_8 \;,
\eqa
which can be conveniently cast in a form that makes the axial rotation of the normal vacuum transparent~\cite{kim} 
$\Sigma_{\alpha}=A_{\alpha}\Sigma_0A_{\alpha}$,
\bqa
A_{\alpha}
&=&{1+2\cos{\frac{\alpha}{2}}\over3}\mathbb{1}+i\lambda_2\sin\frac{\alpha}{2}
+{\cos\frac{\alpha}{2}-1\over\sqrt{3}}\lambda_8 
\eqa
and
$\Sigma_0=\mathbb{1}$. 
As pointed out in Ref.~\cite{kim} and
discussed in some detail in Ref.~\cite{usagain}, parameterizing fluctuations around the rotated vacuum requires an equivalent rotation of the generators, without which the theory is not renormalizable and the kinetic terms are non-canonical. The upshot is that the $\Sigma$ fields in the $\chi$PT Lagrangian should be written as
\bqa
\Sigma&=&L_{\alpha}\Sigma_{\alpha}R_{\alpha}^{\dagger}\;,
\label{para}
\eqa
where
\bqa
\label{la}
L_{\alpha}&=&A_{\alpha}UA_{\alpha}^{\dagger}\\
R_{\alpha}&=&A_{\alpha}^{\dagger}U^{\dagger}A_{\alpha}\;,
\label{ra}
\eqa
which guarantees that the fluctuations around the rotated ground state are parameterized correctly. $U$ is defined as
\bqa
U&=&e^{i{\phi_i\lambda_i\over2f}}\;,
\eqa
where $\phi_{i}$ are the fluctuations of the pion, kaon and eta fields and $\lambda_{i}$ are the unrotated generators. It follows from the parametrization above that
\bqa
\Sigma&=&A_{\alpha}(U\Sigma_0U)A_{\alpha}
=A_{\alpha}U^2A_{\alpha}\;,
\eqa
where $\Sigma=U^2$ when $\alpha=0$.

Expanding the leading order $\chi$PT Lagrangian using $\Sigma$ we get the following structure
\bqa
{\cal L}_2&=&{\cal L}_2^{\rm static}
+{\cal L}_2^{\rm linear}+{\cal L}_2^{\rm quadratic}+...\;,
\eqa
where ${\cal L}_2^{\rm static}$ is the contribution with no derivatives or fluctuations, ${\cal L}_2^{\rm linear}$ is linear in the fields and ${\cal L}_2^{\rm quadratic}$ is quadratic. Explicitly,
\bqa
\nonumber
\label{lagstat}
{\cal L}_2^{\rm static}&=&f^2B_0(2m_j+m_s)\\
&&+{1\over2}f^2(\mu_I\sin\alpha+a_{0}^{1}\cos\alpha)^{2}\;,\\ \nonumber
{\cal L}_2^{\rm linear}&=&-2fB_0\bar{m}_j\phi_{2}\\
\nonumber
&&+f(\mu_I\sin\alpha+a_{0}^{1}\cos\alpha)\mu_{I}\cos\alpha\phi_2\\
\nonumber
&&-f(\mu_I\sin\alpha+a_0^1\cos\alpha)a_0^1\sin\alpha\phi_2 \\
&&-f(\mu_I\sin\alpha+{a_0^1\cos\alpha)}\partial_0
\phi_{1}\;,
\label{lini}
\\ \nonumber
  \mathcal{L}_{2}^{\rm quadratic}&=&\frac{1}{2}\partial_{\mu}\phi_{a}
  \partial^{\mu}\phi_{a}-{1\over2}m_a^2\phi_a^2\\ \nonumber
  &&+\frac{1}{2}m_{12}(\phi_{1}\partial_{0}\phi_{2}-\phi_{2}\partial_{0}\phi_{1})\\
  \nonumber
&&+\frac{1}{2}m_{45}\left (\phi_{4}\partial_{0}\phi_{5}-\phi_{5}\partial_{0}\phi_{4}\right )\\
&&+\frac{1}{2}m_{67}\left (\phi_{7}\partial_{0}\phi_{6}-\phi_{6}\partial_{0}\phi_{7}\right),\;
\eqa
where we first define $j$-dependent masses in order to make the notation leaner
\bqa
m_j&=&m\cos\alpha+j\sin\alpha\;,\\
\bar{m}_j&=&m\sin\alpha-j\cos\alpha\;.
\eqa
The masses in the Lagrangian in terms of $m_{j}$ and $\bar{m}_{j}$ in the pion sector are
\bqa
m_1^2&=&2B_{0}m_{j}{-(\mu_{I}
\cos\alpha-a_0^1\sin\alpha)^2},\\ \nonumber
m_2^2&=&
2B_{0}m_{j}
-(\mu_I\cos\alpha-a_0^1\sin\alpha)^2
\\ 
&+&(\mu_I\sin\alpha+a_0^1\cos\alpha)^2\;,
\\
m_{12}&=&2(\mu_{I}\cos\alpha-a_{0}^{1}\sin\alpha)\;,\\
m_3^2&=&2B_{0}m_{j}+(\mu_I\sin\alpha+a_0^1\cos\alpha)^2\;,
\eqa
in the charged kaon sector are
\bqa
\nonumber
m_{4}^{2}&=&B_{0}m_{s}+B_{0}m_{j}+\frac{1}{4}(\mu_I\sin\alpha+a_0^1\cos\alpha)^2\\
&&-\frac{1}{4}(\mu_I\cos\alpha-a_0^1\sin\alpha)^2\;,\\
m_{5}^{2}&=&m_{4}^{2}\;,\\
m_{45}&=&(\mu_{I}\cos\alpha-a_{0}^{1}\sin\alpha)\;.
\eqa
in the neutral kaon sector are
\bqa
m_{6}^{2}&=&m_{4}^{2}\;,\\
m_{7}^{2}&=&m_{4}^{2}\;,\\
m_{67}&=&m_{45}\;,
\eqa
and finally the eta mass is
\bqa
m_{8}^{2}&=&\frac{2B_{0}}{3}({2}m_{s}+m_{j})\;.
\eqa
Since in the following sections we will need these masses in the limit of a zero axial vector source, we adopt the following equals sign convention whereby any equation that follows $\overset{a=0}{=}$ is assumed to be in this limit. For instance,
\begin{equation}
    m_{1}^{2}\overset{a=0}{=}2B_{0}m_{j}-\mu_{I}^{2}\cos^2\alpha\;.
\end{equation}
Next, using the quadratic Lagrangian, we find the inverse
propagator:
\bqa
D^{-1}&=&
\begin{pmatrix}
D^{-1}_{12}&0&0&0&0\\
0&P^{2}-m_{3}^{2}&0&0&0\\
0&0&D^{-1}_{45}&0&0\\
0&0&0&D^{-1}_{67}&0\\
0&0&0&0&P^{2}-m_{8}^{2}\\
\end{pmatrix}\;,
\eqa
where $P=(p_0,p)$ is the four-momentum in Minkowski space, such that $P^2=p_0^2-p^2$. The inverse propagator for the charged pions is $D^{-12}$, the charged kaons is $D^{-1}_{45}$ and the neutral kaons is $D^{-1}_{67}$
\begin{align}
D^{-1}_{12}&=
\begin{pmatrix}
P^{2}-m_{1}^{2}&ip_{0}m_{12}\\
-ip_{0}m_{12}&P^{2}-m_{2}^{2}\\
\end{pmatrix}\;,
\\  
D^{-1}_{45}&=
\begin{pmatrix}
P^{2}-m_{4}^{2}&ip_{0}m_{45}\\
-ip_{0}m_{45}&P^{2}-m_{5}^{2}\\
\end{pmatrix}\;,
  \\
D^{-1}_{67}&=
\begin{pmatrix}
P^{2}-m_{6}^{2}&ip_{0}m_{67}\\
-ip_{0}m_{67}&P^{2}-m_{7}^{2}\\
\end{pmatrix}\;,
\end{align}
with the masses defined above. In order to renormalize the one-loop effective potential we also need the tree-level contribution from the $\mathcal{O}(p^{4})$ $\chi$PT Lagrangian~\cite{gasser2}, where the relevant terms are 
\bqa\nonumber
{\cal L}_4&=&
 L_1\left({\rm Tr}
\left[\nabla_{\mu}\Sigma^{\dagger}\nabla^{\mu}\Sigma\right]\right)^2
\\ && \nonumber+L_2{\rm Tr}\left[\nabla_{\mu}\Sigma^{\dagger}\nabla_{\nu}\Sigma\right]
    {\rm Tr}\left[\nabla^{\mu}\Sigma^{\dagger}\nabla^{\nu}\Sigma\right]
\\ && \nonumber
    +L_3{\rm Tr}\left[(\nabla_{\mu}\Sigma^{\dagger}\nabla^{\mu}\Sigma)
(\nabla_{\nu}\Sigma^{\dagger}\nabla^{\nu}\Sigma)\right]
\\&& \nonumber
+L_4{\rm Tr}\left[\nabla_{\mu}\Sigma^{\dagger}\nabla^{\mu}\Sigma\right]
{\rm Tr}\left[\chi^{\dagger}\Sigma+\chi\Sigma^{\dagger}\right]
\\ &&\nonumber
+L_5{\rm Tr}\left[\left(\nabla_{\mu}\Sigma^{\dagger}\nabla^{\mu}\Sigma\right)
\left(\chi^{\dagger}\Sigma+\chi\Sigma^{\dagger}\right)\right]
\\ && \nonumber+L_6\left({\rm Tr}\left[\chi^{\dagger}\Sigma
    +\chi\Sigma^{\dagger}\right]\right)^2
\\ &&
+L_8{\rm Tr}\left[\chi^{\dagger}\Sigma \chi^{\dagger}\Sigma
+  \chi\Sigma^{\dagger}\chi\Sigma^{\dagger}\right]
+H_2{\rm Tr}[\chi\chi^{\dagger}]\;,
\label{lag}
\eqa
where the low energy constants $L_{i}$ and $H_{i}$ are defined as~\cite{gasser2}
\bqa
L_i&=&L_i^r-{\Gamma_i\Lambda^{-2\epsilon}\over2(4\pi)^2}\left[
{1\over\epsilon}+1\right]\;,
\label{renorm1}
\\
H_i&=&L_i^r-{\Delta_i\Lambda^{-2\epsilon}\over2(4\pi)^2}\left[
  {1\over\epsilon}+1\right]\;.
\label{renorm2}
\eqa
The constants
$\Gamma_i$ and $\Delta_i$ assume the following values~\cite{gasser2}
\begin{align}
&  \Gamma_{1}=\frac{3}{32}\;,& \Gamma_{2}&=\frac{3}{16}\;,& \Gamma_{3}&=0\;,
&  \Gamma_{4}={1\over8}\;,
  \\
    &  \Gamma_{5}=\frac{3}{8}\;,& \Gamma_{6}&=\frac{11}{144}\;,& 
\Gamma_{8}&={5\over48}\;,
&  \Delta_{2}={5\over24}\;.
\end{align}
$L_{i}^{r}$ and $H_{i}^{r}$ are scale-dependent and run in order to ensure the scale independence of physical quantities observables in $\chi$PT, as follows
\bqa
\Lambda{dL_i^r\over d\Lambda}=-{\Gamma_i\over(4\pi)^2}\;,
\hspace{1cm}
\Lambda{dH_i^r\over d\Lambda}=-{\Delta_i\over(4\pi)^2}\;.
\eqa
We only need the static contribution from $\mathcal{L}_{4}$, Eq.~(\ref{lag}), 
which is given below,
\bqa\nonumber
{\cal L}_4^{\rm static}&=&(4L_1+4L_2+2L_3)(\mu_I\sin\alpha+a_{0}^{1}\cos\alpha)^{4}
\\
\nonumber
&+&8L_4B_0(2m_j+m_s)(\mu_I\sin\alpha+a_{0}^{1}\cos\alpha)^{2}\\
\nonumber
&+&8L_5B_0m_j(\mu_I\sin\alpha+a_{0}^{1}\cos\alpha)^{2}\\
\nonumber
&+&16L_6B_0^2(2m_j+m_s)^2\\
\nonumber
&+&8L_8B_0^2\left[2m_j^2-2\bar{m}_j^2+m_s^2\right]\\
&+&4H_2B_0^2\left[2m_j^2+2\bar{m}_j^2+m_s^2\right]\;.
\label{stat4}
\eqa

\section{Effective potential}  
\label{Veffective}
In this section, we calculate the next-to-leading order
effective potential using the Lagrangian from the previous section. We begin with the tree-level effective potential $V_{0}$,
which is simply given by $V_0=-{\cal L}_{2}^{\rm static}$,
\bqa\nonumber
V_0&=&-f^2B_0
(2m_j+m_s)
-{1\over2}{f^2}
(\mu_I\sin\alpha+a_{0}^{1}\cos\alpha)^{2}\ .
\\ &&
\label{v0} 
\eqa
Similarly, the next-to-leading order static contribution $V_1^{\rm static}$ is given by $V_1^{\rm static}=-{\cal L}_4^{\rm static}$,
\bqa\nonumber
V_1^{\rm static}&=&-(4L_1+4L_2+2L_3)(\mu_I\sin\alpha+a_{0}^{1}\cos\alpha)^{4}\\
\nonumber
&-&8L_4B_0(2m_j+m_s)(\mu_I\sin\alpha+a_{0}^{1}\cos\alpha)^{2}\\
\nonumber
&-&8L_5B_0m_j(\mu_I\sin\alpha+a_{0}^{1}\cos\alpha)^{2}\\
\nonumber
&-&16L_6B_0^2(2m_j+m_s)^2\\
\nonumber
&-&8L_8B_0^2\left[2m_j^2-2\bar{m}_j^2+m_s^2\right]\\
&-&4H_2B_0^2\left[2m_j^2+2\bar{m}_j^2+m_s^2\right]\;.
\label{v1stat}
\eqa
The one-loop contributions from the neutral pion and the eta meson are of the form
\bqa
V_1&=&\frac{1}{2}\int_P\log\left[P^2+m^2\right]\;.
\label{logint}
\eqa
The integral can be evaluated easily using dimensional regularization and the result is stated in Appendix A, Eq.~(\ref{sumint0}).
The one-loop contribution from the charged pions on the other hand is of the form
\bqa\nonumber
V_{1,\pi^+}+V_{1,\pi^-}&=&
{1\over2}\int_P\log[(p_0^2+E_{\pi^+}^2)(p_0^2+E_{\pi^-}^2)]
\\ &=&
{1\over2}\int_p\left[{E_{\pi^+}}+{E_{\pi^-}}\right]\;,
\eqa
where the energies $E_{\pi^{\pm}}$ are given by
\bqa
\nonumber
E_{\pi^{\pm}}^{2}&=&p^2+{1\over2}\left(m_{1}^{2}+m_{2}^{2}+m_{12}^{2}\right)
\\ 
&&
\hspace{-0.5cm}
\pm{1\over2}\sqrt{4p^{2}m_{12}^{2}+(m_{1}^{2}+m_{2}^{2}
  +m_{12}^{2})^2-4m_{1}^{2}m_{2}^{2}}\;.
\label{pipo}
\eqa
In order to isolate the divergences, we expand 
$E_{\pi^{\pm}}$
in powers of $p$ 
around infinity up to terms that contain divergences
\bqa\nonumber
E_{\pi^+}+E_{\pi^-}&=&
2p+\frac{2(m_{1}^{2}+m_{2}^{2})+m_{12}^{2}}{4p}
\\ &&
\hspace{-1.5cm}
-\frac{8(m_{1}^{4}+m_{2}^{4})+4(m_{1}^{2}+m_{2}^{2})m_{12}^{2}
  +m_{12}^{4}}{64p^{3}}+...
\label{expandp}
\eqa
Noting that the divergences in 
Eq.~(\ref{expandp}) are the same as those of
$E_1+E_2$, where
$E_{1,2}=\sqrt{p^2+m_{1,2}^2+\mbox{$1\over4$}m_{12}^2}=\sqrt{p^2+\tilde{m}^2_{1,2}}$,
$\tilde{m}_1^2=2B_0m_j$, $\tilde{m}_2^2=m_3^2$,  we can
isolate the divergences by  writing
\bqa
\label{divint}
V_{{\rm 1},\pi^{+}}^{\rm div}+V_{{\rm 1},\pi^{-}}^{\rm div}
&=&{1\over2}\int_p\left[E_1+E_2\right]
\;,\\
V_{{\rm 1},\pi^{+}}^{\rm fin}+V_{{\rm 1},\pi^{-}}^{\rm fin}
&=&\frac{1}{2}\int_{p}\left [E_{\pi^+}+E_{\pi^-}-E_1-E_2\right ]\;,
\label{subt}
\eqa
where the divergent and finite parts of the charged pion integrals satisfy
\bqa
V_{1,\pi^+}+V_{1,\pi^-}
&=&V_{{\rm 1},\pi^{+}}^{\rm div}+V_{{\rm 1},\pi^{-}}^{\rm div}
+
V_{{\rm 1},\pi^{+}}^{\rm fin}+V_{{\rm 1},\pi^{-}}^{\rm fin}\;.
\eqa
The one-loop contribution to the effective potential from the charged kaons is
\bqa
V_{1,K^{+}}+V_{1,K^{-}}
&=&{1\over2}\int_P\log\left[
  \left(P^2+m_4^2\right)^2+p_0^2m_{45}^2
\right],\;
\eqa
where we have used that $m_4=m_5$. The integrand  can be factorized and the
integral rewritten as
\bqa\nonumber
V_{1,K^{+}}+V_{1,K^{-}}
&=&{1\over2}\int_P\log\left\{
\left[\left(p_0+\mbox{$1\over2$}{im_{45}}\right)^2+p^2
+\tilde{m}_4^2
\right]
\right.\\ &&\times\left.
\left[\left(p_0-\mbox{$1\over2$}im_{45}\right)^2+p^2
  +\tilde{m}_4^2
\right]
\right\}
\;,
\eqa
with $\tilde{m}_4^2=m_4^2+{1\over4}m_{45}^2$. Replacing the terms in the parenthesis as
$p_0\pm{im_{45}\over2}\rightarrow p_0$, which is permitted since the $p_{0}$ integral is being performed from negative infinity to positive infinity, we get a simple expression for the one-loop contribution from the charged kaons in terms of $\tilde{m}_{4}$,
\bqa
V_{1,K^{+}}+V_{1,K^{-}}
&=&\int_P\log\left[P^2+\tilde{m}_4^2\right]\;.
\label{kaonkut}
\eqa
Noting that the contribution from the neutral kaons is identical since $\log D_{67}^{-1}=\log D_{45}^{-1}$, using Eq.~(\ref{kaonkut}) and Eq.~(\ref{sumint0}), we get for the divergent contribution to the full one-loop potential
\bqa\nonumber
V_{\rm 1}^{\rm div}
&=&
-\frac{\tilde{m}_1^4}{4(4\pi)^{2}}\left [\frac{1}{\epsilon}+\frac{3}{2} +
  \log\left (\frac{\Lambda^{2}}{{\tilde{m}_1^2}}
  \right ) \right ]
\\ &&\nonumber-\frac{\tilde{m}_2^4}{4(4\pi)^{2}}
\left [\frac{1}{\epsilon}+\frac{3}{2} +
  \log\left (\frac{\Lambda^{2}}{{\tilde{m}_2^2}}
  \right ) \right ]
\\ &&\nonumber
-\frac{m_3^4}{4(4\pi)^{2}}\left [\frac{1}{\epsilon}+\frac{3}{2} +
  \log\left (\frac{\Lambda^{2}}{m_{3}^{2}} \right )\right ]
\\ \nonumber
&&-\frac{\tilde{m}_4^4}{(4\pi)^{2}}\left [\frac{1}{\epsilon}+\frac{3}{2} +
  \log\left (\frac{\Lambda^{2}}{\tilde{m}_4^{2}} \right )\right ]
\\ 
&& 
-\frac{m_8^4}{4(4\pi)^{2}}\left [\frac{1}{\epsilon}+\frac{3}{2} +
  \log\left (\frac{\Lambda^{2}}{m_{8}^{2}} \right )\right ]
\label{divpi}
\;.
\eqa
Combining Eq.~(\ref{divpi}) with the tree-level 
contribution from $\mathcal{L}_{2}$, the counterterm
$\mathcal{L}_{4}$, and renormalization of the couplings $L_i$ and $H_i$
according to Eqs.~(\ref{renorm1})--(\ref{renorm2}),
we get the final form of the one-loop effective potential
\bqa\nonumber
V_{\rm eff}&=&-f^2B_0(2m_j+m_s)
-{1\over2}f^2(\mu_I\sin\alpha+a_{0}^{1}\cos\alpha)^{2}
\\ && \nonumber
-\left[
  64L_6^r+16L_8^r+8H_2^r
     +{1\over(4\pi)^2}\left({37\over18}
+\log{\Lambda^2\over\tilde{m}_1^2}
\right.\right.  \\&& \left.\left.\nonumber
  +2\log{\Lambda^2\over m_3^2}+
    \log{\Lambda^2\over\tilde{m}_4^2}
+{1\over9}\log{\Lambda^2\over m_8^2}
  \right)\right]B_0^2m_j^2
\\ && \nonumber
-\left[  64L_6^r
  +{1\over(4\pi)^2}\left({11\over9}+
    2\log{\Lambda^2\over\tilde{m}_4^2}
        +{4\over9}\log{\Lambda^2\over m_8^2}
  \right)\right]
\\ &&\nonumber \times
B_0^2m_jm_s
\\ && \nonumber
-\left[
  16L_6^r+8L_8^r+4H_2^r
  +{1\over(4\pi)^2}\left({13\over18}+\log{\Lambda^2\over\tilde{m}_4^2}
\right.\right.  \\&& \left.\left.\nonumber
  +{4\over9}\log{\Lambda^2\over m_8^2}\right)
\right]B_0^2m_s^2
\\ && \nonumber
-\left[8{L}_4^{r}
  +{1\over2(4\pi)^2}\left({1\over2}+\log{\Lambda^2\over\tilde{m}_4^2}\right)
\right]
\\ &&\nonumber\times
                  B_0(2m_j+m_s)(\mu_I\sin\alpha+a_{0}^{1}\cos\alpha)^{2}
\\ && \nonumber
-\left[8{L}_5^{r}
  +{1\over2(4\pi)^2}\left({3\over2}+4\log{\Lambda^2\over m_3^2}
 -\log{\Lambda^2\over\tilde{m}_4^2}\right)
     \right]
\\ &&\nonumber\times
     B_0m_j(\mu_I\sin\alpha+a_{0}^{1}\cos\alpha)^{2}
+\left(16L_8^r-8H_2^r\right)B_0^2\bar{m}_j^2
\\ &&\nonumber
-\left[4{L}_1^r+4{L}_2^r+2{L}_3^r
  +{1\over16(4\pi)^2}\left({9\over2}+8\log{\Lambda^2\over m_3^2}
\right.\right.  \\&& \left.\left. \nonumber
    +\log{\Lambda^2\over\tilde{m}_4^2}\right)
            \right](\mu_I\sin\alpha+a_{0}^{1}\cos\alpha)^{4}\\
&&+V_{\rm 1,{\pi^{+}}}^{\rm fin}
+V_{\rm 1,{\pi^{-}}}^{\rm fin}
     \;.
\label{renpi}
\eqa
For zero pion and axial sources, $j=0$ and $a_{0}^{1}=0$ respectively, Eq.~(\ref{renpi}) reduces to the result of Ref.~\cite{us}.

\section{Quark, pion and axial condensates}
\label{condensates}
In this section, we calculate the light quark, strange, pion and axial condensates. The up-quark and down-quark condensates are equal in the isospin limit, which we denote as $\langle{\bar{\psi}\psi}\rangle$.
The light quark, strange, pion and axial condensates are then defined as
\bqa
\langle\bar{\psi}\psi\rangle&\equiv&\langle\bar{u}u\rangle=\langle\bar{d}d\rangle\overset{a=0}{=}{1\over2}{\partial V_{\rm eff}\over\partial m}\;,\\
\langle\bar{s}s\rangle&\overset{a=0}{=}&{\partial V_{\rm eff}\over\partial m_s}\;,\\
\langle\pi^+\rangle&\overset{a=0}{=}&
{1\over2}{\partial V_{\rm eff}\over\partial j}\;,\\
\langle \bar{\psi}\tfrac{\lambda_{1}}{2}\gamma^{0}\gamma_{5}\psi\rangle
&\overset{a=0}{=}&\frac{\partial V_{\rm eff}}{\partial a_{0}^{1}}\;.
\eqa
Our definition of the light quark  condensate, $\langle\bar{\psi}\psi\rangle=\langle\bar{u}u\rangle=\langle\bar{dd}\rangle$ is different from the definition used in  the  finite  isospin  lattice  QCD  simulation  of Ref.~\cite{gergy3}, where
$\langle\bar{\psi}\psi\rangle=\langle\bar{u}u\rangle+\langle\bar{dd}\rangle$.
This difference explains the extra factor of ${1\over2}$ in the definition of
the quark condensate above. We also define the pion condensate with an extra factor of ${1\over2}$ compared with the lattice work~\cite{gergy3}. Their pionic source $\lambda$ then  corresponds exactly to our source $j$.
At tree level, the quark, pion and axial condensates are
\bqa
\label{q1}
\langle\bar{\psi}\psi\rangle^{\rm tree}&=&-f^2B_0\cos\alpha=\langle\bar{\psi}\psi\rangle_0^{\rm tree}\cos\alpha
\;,\\
\langle\bar{s}s\rangle^{\rm tree}
&=&-f^2B_0=\langle\bar{s}s\rangle_0^{\rm tree}\;,\\
\langle\pi^+\rangle^{\rm tree}&=&-f^2B_0\sin\alpha
=\langle\bar{\psi}\psi\rangle^{\rm tree}_{0}\sin\alpha\;,\\
\langle \bar{\psi}\tfrac{\lambda_{1}}{2}\gamma^{0}\gamma_{5}\psi\rangle^{\rm tree}&=&-f^2\mu_{I}\sin\alpha\cos\alpha\;,
\label{p1}
\eqa
where the tree-level chiral condensate in the normal vacuum is $\langle\bar{\psi}\psi\rangle_0^{\rm tree}=-f^2B_0$.

The NLO light quark
condensate is
\bqa\nonumber
\langle\bar{\psi}\psi\rangle&\overset{a=0}{=}&
-f^2B_0\cos\alpha\left\{  1+
  \bigg[64L_6^r+16L_8^r+8H_2^r
\right.\\ &&\nonumber \left.
  +{1\over(4\pi)^2}\left(
\log{\Lambda^2\over\tilde{m}_1^2}+2\log{\Lambda^2\over m_3^2}
\right.\right.\\ &&\nonumber \left.\left.
+\log{\Lambda^2\over\tilde{m}_4^2}+{1\over9}\log{\Lambda^2\over{m}_8^2}
\right)
\bigg]{B_0m_j\over f^2}
\right. \\ &&\nonumber\left.
+\left[32L_6^r
    +{1\over(4\pi)^2}\left(\log{\Lambda^2\over\tilde{m}_4^2}
      +{2\over9}
      \log{\Lambda^2\over{m}_8^2}\right)
  \right]{B_0m_s\over f^2} 
\right.\\ && \left.\nonumber
  +\left[8L_4^r+4L_5^r
    +{1\over(4\pi)^2}\left(
      \log{\Lambda^2\over{m}_3^2}
      +{1\over4}\log{\Lambda^2\over\tilde{m}_4^2}
    \right)
  \right]
\right.\\ &&\nonumber\left.
  \times{\mu_I^2\sin^2\alpha\over f^2}
\right\}
+\left(16L_8^r-8H_2^r\right)B_0^2\bar{m}_j\sin\alpha
\\ &&
+{{\frac{1}{2}}}\left({\partial V_{\rm 1,\pi^+}^{\rm fin}\over\partial m}
+{\partial V_{\rm 1,\pi^-}^{\rm fin}\over\partial m}\right)\;.
\label{light}
\eqa
Similarly, the NLO strange-quark condensate is
\bqa\nonumber
\langle\bar{s}s\rangle&\overset{a=0}{=}&-f^2B_0\left\{
  1  +\bigg[64L_6^r
    +{1\over(4\pi)^2}\left(2\log{\Lambda^2\over\tilde{m}_4^2}
\right.\right.\\ &&\left.\left. \nonumber
      +{4\over9}
      \log{\Lambda^2\over{m}_8^2}\right)\bigg]{B_0m_j\over f^2}
    +  \left[32L_6^r+16L_8^r+8H_2^r
\right.\right.\\ &&\left.\left. \nonumber
    +{1\over(4\pi)^2}
    \left(2\log{\Lambda^2\over\tilde{m}_4^2}+{8\over9}
      \log{\Lambda^2\over{m}_8^2}\right)
  \right]{B_0m_s\over f^2}
\right.\\ &&\left.
  +\left[8L_4^r+{1\over2(4\pi)^2}\log{\Lambda^2\over\tilde{m}_4^2}
  \right]{\mu_I^2\sin^2\alpha\over f^2}
\right\}
\;,
\label{sss}
\eqa
where there is no finite effective potential contribution since the kaon one-loop contributions can be written in the standard quadratic form. The result reduces to that of Ref.~\cite{gasser2} in the limit of zero isospin chemical potential.

Using the definition of the pion condensate, we obtain
\bqa\nonumber
\langle\pi^+\rangle&\overset{a=0}{=}&
-f^2B_0\sin\alpha\left\{  1+
  \bigg[64L_6^r+16L_8^r+8H_2^r
\right.\\ &&\left. \nonumber
  +{1\over(4\pi)^2}\left(
\log{\Lambda^2\over\tilde{m}_1^2}+2\log{\Lambda^2\over m_3^2}
+\log{\Lambda^2\over\tilde{m}_4^2}
\right.\right.\\ &&\left.\left. \nonumber
+{1\over9}\log{\Lambda^2\over{m}_8^2}
\right)\bigg]
{B_0m_j\over f^2}
\right. \\ &&\nonumber\left.
+\left[32L_6^r
    +{1\over(4\pi)^2}\left(\log{\Lambda^2\over\tilde{m}_4^2}
      +{2\over9}
      \log{\Lambda^2\over{m}_8^2}\right)
  \right]{B_0m_s\over f^2} 
\right. \\ &&\nonumber\left.
  +\left[8L_4^r+4L_5^r
    +{1\over(4\pi)^2}\left(
      \log{\Lambda^2\over{m}_3^2}
      +{1\over4}\log{\Lambda^2\over\tilde{m}_4^2}
    \right)  \right]
\right.\\ &&\left. \nonumber
  \times{\mu_I^2\sin^2\alpha\over f^2}
\right\}
-\left(16L_8^r-8H_2^r\right)B_0^2\bar{m}_j\cos\alpha
\\ &&
+{{\frac{1}{2}}}\left({\partial V_{\rm 1,\pi^+}^{\rm fin}\over\partial j}
+{\partial V_{\rm 1,\pi^-}^{\rm fin}\over\partial j}\right)\;,
\label{pioncon}
\eqa		
where the condensate vanishes for $\mu_{I}\le m_{\pi}$ since $\alpha=0$.
Finally, the axial condensate, which is zero in the normal vacuum becomes nonzero in the pion condensed phase. The final result is
\bqa\nonumber
\label{axialcon}
\langle{\bar{\psi}\tfrac{\lambda_{1}}{2}\gamma^{0}\gamma_{5}\psi}\rangle&=&
-f^2\mu_I\sin\alpha\cos\alpha\left\{1+\frac{\mu_I^2\sin^2\alpha}{f^2}
\right.\\ && \nonumber
\left[16(L^{r}_{1}+L^{r}_{2})+8L^{r}_{3}
+\frac{1}{4(4\pi)^{2}}\log\frac{\Lambda^{2}}{\tilde{m}_{4}^{2}}
\right.\\ &&\left.\nonumber
+\frac{2}{(4\pi)^{2}}\log\frac{\Lambda^{2}}{m_3^2}\right]
\left.
+\frac{B_0m_j}{f^2}\left[16(2L^{r}_{4}+L^{r}_{5})
\right. \right.
\\ && \nonumber
\left.\left.
+\frac{4}{(4\pi)^{2}}\log\frac{\Lambda^{2}}{m_3^2}+\frac{1}{(4\pi)^{2}}\log\frac{\Lambda^{2}}{\tilde{m}_{4}^{2}}\right]
\right.\\ && \nonumber
\left.
+\frac{B_0m_s}{f^2}\left[
16L^{r}_{4}+\frac{1}{(4\pi)^{2}}\log\frac{\Lambda^{2}}{\tilde{m}_{4}^{2}} \right ]\right\}
\\ &&+
\frac{\partial V_{1,\pi^{+}}^{\rm fin}}{\partial a_0^1}
+\frac{\partial V_{1,\pi^{-}}^{\rm fin}}{\partial a_0^1}\;,
\eqa
where setting $\alpha=0$ gives zero as required since pion condensation is required for the axial condensate to form.

\subsection{Two-flavor $\chi$PT in the large-$m_{s}$ limit}
\label{mapping}
In the limit $m_s\gg m_u=m_d$, we expect 
using effective field theory arguments that
the degrees of freedom
containing an $s$-quark, i.e. the kaons and the eta, to decouple.
Our results for the light-quark condensate and the pion condensate should
then reduce to the two-flavor case, albeit with renormalized
couplings. The only reference left to the $s$-quark is in the
expressions for the modified couplings
$l_i^r$ and $h_i^r$, and modified parameters $\tilde{f}$ and $\tilde{B}$, 
see Eqs.~(\ref{run1})--(\ref{rel6}) below.
This was shown explicitly in Ref.~\cite{gasser2}, where relations among the low-energy constants in two - and three-flavor $\chi$PT were derived. 

We begin by expanding the light-quark condensate in inverse powers of $m_s$. Eq.~(\ref{light}) then reduces to
\bqa\nonumber
\langle\bar{\psi}\psi\rangle&=&-\tilde{f}^2\tilde{B}_0
\cos\alpha\left\{
  1+  \bigg[4l_3^r+4l_4^r
\right.\\ &&\left. \nonumber
  +{1\over(4\pi)^2}\left(
\log{\Lambda^2\over\tilde{m}_1^2}+2\log{\Lambda^2\over m_3^2}
\right)
\bigg]{B_0m_j\over f^2}
\right. \\ &&\nonumber\left.
  +\left[l_4^r
    +{1\over(4\pi)^2}
      \log{\Lambda^2\over{m}_3^2}
      \right]
  {\mu_I^2\sin^2\alpha\over f^2}
\right\}
\\ &&
+4({l}_4^r-{h}_1^r)B_0^2m
+\frac{1}{2}\left({\partial V_{\rm 1,\pi^+}^{\rm fin}\over\partial m}
+{\partial V_{\rm 1,\pi^-}^{\rm fin}\over\partial m}\right)\;,
\label{light2}
\eqa
where we have introduced new renormalized couplings 
$l_1^r$--$l_4^r$, and $h_1^r$, as well as modified parameters $\tilde{f}$ and $\tilde{B}_0$, which we define below,
\bqa\nonumber
\label{start1}
l_1^r+l_2^r&=&4(L_1^r+L_2^r)+2L_3^r+\frac{1}{16(4\pi)^2}\left[
  \log{\Lambda^2\over \tilde{m}_{K,\textrm{0}}^2}-1\right]
  \\
  && \\
l_3^r+l_4^r&=&
16L_6^r+8L_8^r
+{1\over4}{1\over(4\pi)^2}\left[
  \log{\Lambda^2\over \tilde{m}_{K,\textrm{0}}^2}-1\right]
\\ && 
\label{run1}
+{1\over36}{1\over(4\pi)^2}
\left[\log{\Lambda^2\over \tilde{m}_{\eta,\textrm{0}}^2}-1\right]
\;,\\ 
l_4^r&=&8L_4^r+4L_5^r
+{1\over4}{1\over(4\pi)^2}\left[
  \log{\Lambda^2\over \tilde{m}_{K,\textrm{0}}^2}-1\right]\;,
\label{run2}
\\
l_4^r-h_1^r&=&4L_8^r-2H_2^r\;,
\label{rel4}\\ \nonumber
\tilde{f}^2&=&f^2\left[1+
  \left(16L_4^r
    +{1\over(4\pi)^2}\log{\Lambda^2\over \tilde{m}_{K,0}^2}\right)
{B_0m_s\over f^2}\right]\;,
\\ &&
\label{rel5}
\\ \nonumber
\tilde{B}_0&=&B_0\left.\Bigg[
  1-  \Bigg (16L_4^r-32L_6^r
\right.\Bigg. \\ && \Bigg.\left.
-    {2\over9(4\pi)^2}\log{\Lambda^2\over \tilde{m}_{\eta,0}^2}\right){B_0m_s\over f^2}
\Bigg]\;.
\label{rel6}
\eqa
The new mass parameters are defined as $\tilde{m}_{K,0}^2=B_0m_s$
and $\tilde{m}_{\eta,0}^2={4B_0m_s\over3}$.
The parameters $\tilde{B}_0$ and $\tilde{f}$ can also be obtained by
considering the one-loop expressions for the chiral condensate and the
pion decay constant ignoring the loop corrections from the pions, i.e. 
they are obtained by integrating out the $s$-quark.

The relations between the renormalized couplings $l_i^r$ and the
low-energy constants $\bar{l}_i$ in two-flavor $\chi$PT are
\bqa
\label{2fLECl}
l_i^r(\Lambda)&=&{\gamma_i\over2(4\pi)^2}
\left[\bar{l}_i+\log{2B_0m\over\Lambda^2}\right]\;,
\\
h_i^r(\Lambda)&=&{\delta_i\over2(4\pi)^2}
\left[\bar{h}_i+\log{2B_0m\over\Lambda^2}\right]\;,
\label{2fLEC}
\eqa
where $\gamma_1={1\over3}$, $\gamma_2={2\over3}$, $\gamma_3=-{1\over2}$, 
$\gamma_4=2$, and $\delta_1=2$~\cite{gasser1}.
These equations can be used to calculate the running of the
couplings $l_i^r$ and $h_i^r$ with the renormalization scale.
One can then verify that the running of the left-hand and right-hand side
of Eqs.~(\ref{run1})--(\ref{run2}) is the same. One can also verify
that the modified parameters $\tilde{f}^2$ and $\tilde{B}_0$ do not run.
Inserting these relations into Eq.~(\ref{light}), we find
\bqa\nonumber
\langle\bar{\psi}\psi\rangle&=&-\tilde{f}^2\tilde{B}_0
\cos\alpha\left[1+  {1\over(4\pi)^2}
  \left(4\bar{l}_4-\bar{l}_3
 \right.\right.\\ && \nonumber\left.\left.
   +  \log{2B_0m\over\tilde{m}_1^2}
   +2\log{2B_0m\over m_3^2}
       \right){B_0m_j\over f^2}
\right.\\ &&\nonumber \left.
  +{1\over(4\pi)^2}\left(\bar{l}_4+\log{2B_0m\over m_3^2}\right)
  {\mu_I^2\sin^2\alpha\over f^2}
\right]
\\ &&
+{4\tilde{B}_0^2{m}\over(4\pi)^2}\left(\bar{l}_4-\bar{h}_1\right)
+\frac{1}{2}\left({\partial V_{\rm 1,\pi^+}^{\rm fin}\over\partial m}
+{\partial V_{\rm 1,\pi^-}^{\rm fin}\over\partial m}\right).\;
\label{finquark}
\eqa
The pion condensate can be calculated in the same way and the result is
\bqa\nonumber
\langle\pi^+\rangle&=&-\tilde{f}^2\tilde{B}_0
\sin\alpha\left[1+  {1\over(4\pi)^2}
  \left(4\bar{l}_4-\bar{l}_3
  \right.\right.\\ &&\left.\left.\nonumber
    +\log{2B_0m\over\tilde{m}_1^2}
    +2\log{2B_0m\over m_3^2} 
  \right){B_0m_j\over f^2}
\right.\\ &&\nonumber \left.
  +{1\over(4\pi)^2}\left(\bar{l}_4+\log{2B_0m\over m_3^2}\right)
    {\mu_I^2\sin^2\alpha\over f^2}
  \right]
\\  && 
+{4\tilde{B}_0^2{j}\over(4\pi)^2}\left(\bar{l}_4-\bar{h}_1\right)
+\frac{1}{2}\left({\partial V_{\rm 1,\pi^+}^{\rm fin}\over\partial j}
+{\partial V_{\rm 1,\pi^-}^{\rm fin}\over\partial j}\right).\;
\eqa
Finally, the axial condensate in the large-$m_s$ limit is
\begin{align}
    &\langle{\bar{\psi}\tfrac{\tau_{1}}{2}\gamma^{0}\gamma_{5}\psi}\rangle\overset{a=0}{=}-\tilde{f}^2\mu_I\sin\alpha\cos\alpha\left\{1+\frac{\mu_I^2\sin^2\alpha}{(4\pi f)^2}\right. \nonumber \\
    &\left.\times \left[\frac{2}{3}\left(\bar{l}_1+2\bar{l}_2\right)+2\log\frac{2B_0m}{m_3^2}\right]+\frac{\tilde{B}_0m_j}{(4\pi f)^2}\left[4\bar{l}_4\right.\right. \nonumber \\
    &\left.\left. +4\log\frac{2B_0m}{m_3^2}\right]\right\}+\frac{\partial V^{\rm fin}_{1,\pi^+}}{\partial a_{0}^{1}}+\frac{\partial V^{\rm fin}_{1,\pi^-}}{\partial a_{0}^{1}}\;.
\end{align}
In order to evaluate the condensates in two-flavor $\chi$PT, we need the ground state value of $\alpha$, which is obtained from the effective potential of two-flavor $\chi$PT in the presence of a pseudo-scalar source. The two-flavor effective potential can be found by taking the large-$m_{s}$ limit in the effective potential Eq.~(\ref{renpi}) and the identification of two-flavor LECs as was done with the condensates. We obtain
\bqa\nonumber
V_{\rm eff}&=&-2\tilde{f}^{2}\tilde{B}_{0}m_{j}-\frac{1}{2}\tilde{f}^{2}(\mu_{I}\sin\alpha+a_{0}^{2}\cos\alpha)^{2}\\ \nonumber
&&-\frac{1}{(4\pi)^{2}}\left[\frac{3}{2}-\bar{l}_{3}+4\bar{l}_{4}+\log\left(\frac{2B_{0}m}{{\tilde{m}_1^{2}}}\right)\right.\\ \nonumber
&&\left.+2\log\left(\frac{2B_{0}m}{m_{3}^{2}}\right)\right ]B_{0}^{2}m_{j}^{2}\\
\nonumber
&&-\frac{1}{(4\pi)^{2}}\left[\frac{1}{2}+\bar{l}_{4}+\log\left(\frac{2B_{0}m}{m_{3}^{2}}\right) \right]2B_{0}m_{j}\\ \nonumber
&&\times(\mu_{I}\sin\alpha+a_{0}^{2}\cos\alpha)^{2}\\ \nonumber
&&-\frac{1}{2(4\pi)^{2}}\left[\frac{1}{2}+\frac{1}{3}\bar{l}_{1}+\frac{2}{3}\bar{l}_{2}+\log\left(\frac{2B_{0}m}{m_{3}^{2}}\right) \right ]\\ \nonumber
&&\times(\mu_{I}\sin\alpha+a_{0}^{2}\cos\alpha)^{4}\\  
\nonumber
&&+\frac{4}{(4\pi)^{2}}\left (\bar{l}_{4}-\bar{h}_{1} \right)B_0^2\left[m_j^2+\bar{m}_j^2
\right]\\ 
&&+V^{\rm fin}_{1,\pi^{+}}+V^{\rm fin}_{1,\pi^{-}}\;.
\label{vlargems}
\eqa
Taking appropriate derivatives of the two-flavor effective potential 
Eq.~(\ref{vlargems})
yields the various condensates. However, note that $2B_0m$ is a reference scale $M$ and must be held
fixed when taking the partial derivative with respect to $m$ to obtain the quark condensate.
We note that in the two-flavor effective potential and the condensates, $\tilde{B}_{0}$ of Eq.~(\ref{rel6}) appears in the leading order terms and $B_{0}$ appears in the next-to-leading order terms. The large-$m_{s}$ limit we perform has the following formal ordering of the various scales,
\begin{equation}
\begin{split}
B_{0}m_{u}=B_{0}m_{d}\ll B_{0}m_{s}\ll (4\pi f_{\pi})^{2}\ .
\end{split}
\end{equation}
The first equality is the isospin limit, the second the large-$m_{s}$ limit, and the last ensures the validity of an effective field theory approach. In
this formal limit the $B_{0}$ in the next-to-leading order result can be identified with $\tilde{B}_{0}$ up to the order we are working. Finally, our expansion in inverse powers of $m_s$ also assumes $B_0j\ll B_0m_s$ and $\mu_I^2\ll B_0m_s$.

\section{Numerical results and discussion}
\label{numerics}
In this section, we use the results from the previous section to plot the strange-quark condensate and the axial condensate at zero pionic and axial sources. We also plot the light quark and pion condensate 
for nonzero pionic source.
Finally, we compare the nonzero pionic source results with lattice simulations and compare the axial condensate with available lattice results at zero pionic and axial sources.

Finite isospin QCD on the lattice is studied by adding an explicit pionic source since spontaneous symmetry breaking in finite volume is forbidden. Obtaining the pion condensate then requires not just taking the continuum limit but also extrapolating to a zero external source, which is a difficult procedure. We also note that the quark, pion and axial condensates given by Eqs.~(\ref{light})-(\ref{axialcon}) depend on the ground state value of $\alpha$, which can be found by minimizing the one-loop effective potential, i.e. solving
${\partial V_{\rm eff}\over\partial\alpha}=0$, at zero axial vector source.

\subsection{Definitions and choice of parameters}

Since we are interested in the condensates as functions of the isospin chemical potential $\mu_I$, i.e. in medium effects, 
we plot the (normalized) change in the chiral condensate, strange-quark condensate, the pion condensate and the axial condensate relative to the normal vacuum using the following definitions~\cite{gergy3}
\begin{equation}
\begin{split}
\label{deviation}
\Sigma_{\bar{\psi}\psi}&=-\frac{2m}{m_{\pi}^{2}f_{\pi}^{2}}\left[\langle\bar{\psi}\psi\rangle_{\mu_{I}}^{a=0}-\langle\bar{\psi}\psi\rangle_{0}^{a=j=0}\right]+1\;,\\
\Sigma_{\pi}&=-\frac{2m}{m_{\pi}^{2}f_{\pi}^{2}}\langle\pi^{+}\rangle^{a=0}_{\mu_{I}}\;,\\
{\Sigma_{\bar{s}s}}&
=-\frac{m+m_{s}}{m_{K}^{2}f_{K}^{2}}\left[\langle\bar{s}s\rangle^{a=0}_{\mu_{I}}-\langle\bar{s}s\rangle^{a=0}_{0}\right]+1
\;,\\
\Sigma_{a}&=-\langle\bar{\psi}\tfrac{\lambda_1}{2}\gamma^{0}\gamma_{5}\psi\rangle^{j=0}_{\mu_{I}}\;.
\end{split}
\end{equation} 
Note that $\Sigma_{a}$ is simply the negative of the axial condensate and the normalization has been chosen to match that of lattice QCD~\cite{pionstar}. The chiral and pion condensate deviations satisfy
\begin{equation}
\begin{split}
\Sigma_{\bar{\psi}\psi,\rm tree}^{2}+\Sigma_{\pi,\rm tree}^{2}=1\ ,
\label{chiralcircle}
\end{split}
\end{equation}
at tree level in both the normal vacuum and the pion condensate phases even in the presence of a pseudo-scalar source. For the calculations of the deviations and the axial condensate we will use the following values of the quark masses allowing for a $5\%$ uncertainty, consistent with Ref.~\cite{BMW},
\begin{align}
\label{mq}
m_{u}&=2.15
\;{\rm MeV}
,\; m_{d}=4.79
\;{\rm MeV}\;,
\\
m&=\frac{m_{u}+m_{d}}{2}=3.47\;{\rm MeV}\;,\\
B_0m_s&=
m_{K,\rm tree}^{2}-{1\over2}m_{\pi,\rm tree}^{2} ,\;
\end{align}
where $m_{\pi,\rm tree}$ and $m_{K,\rm tree}$ are the tree level pion mass and kaon mass respectively. Note that since $B_{0}$ is fixed by the up and down quark masses and the GOR relation, the strange quark mass is fixed in three-flavor $\chi$PT by the value of $B_{0}$ and the tree level pion and kaon masses. In three-flavor $\chi$PT we cannot fix the strange quark mass independently of the up and down quark masses.
In order to compare with simulations, we adopt the
following values of the pseudo-Nambu-Goldstone masses and decay constants~\cite{private},
\begin{align}
\label{masses}  
m_{\pi}&=131\pm3\;{\rm MeV}\;,
&m_{K}&=481\pm10\;{\rm MeV}\;,\\
f_{\pi}&={128\pm3\over\sqrt{2}}{\rm MeV}\;,
&f_{K}&={150\pm 3\over\sqrt{2}}{\rm MeV}\;.
\end{align}
We point out that the quark masses 
in Eq.~(\ref{mq}) are not those
used in the simulations of Ref.~~\cite{gergy3}
as the latter are unknown.
The quark masses from Ref.~\cite{BMW}
are approximately 3\% larger than given in
Eq.~(\ref{mq}). In Ref.~\cite{2fcondensate},
we therefore varied the quark mass $m_u=m_d$
by 5\% to gauge the sensitivity
of the results. It turns out that the
dominating uncertainty stems from uncertainty of the $\bar{l}_i$s. The same remains true for three-flavor $\chi$PT condensates.

Additionally, we choose the following experimentally determined values for the three-flavor LECs and their associated uncertaintities~\cite{bijnensreview}. The quoted numerical values are at the renormalization scale $\mu=0.77$ GeV, which is approximately the rho-mass, $m_{\rho}$, with $\Lambda^{2}=4\pi e^{-{\gamma}_E}\mu^{2}$~\cite{Hr2ref1,bijnensreview}.
\begin{align}
{L}_{1}^r&=(1.0\pm 0.1)\times10^{-3}\;,\\
{L}_{2}^r&=(1.6\pm 0.2)\times10^{-3}\;,\\
{L}_{3}^r&=(-3.8\pm 0.3)\times10^{-3}\;,\\
{L}_{4}^r&=(0.0 \pm 0.3)\times10^{-3}\;,\\
{L}_{5}^r&=(1.2 \pm 0.1)\times10^{-3}\;,\\
{L}_{6}^r&=(0.0 \pm 0.4)\times10^{-3}\;,\\
{L}_{8}^r&=(0.5 \pm 0.2)\times10^{-3}\;,\\
{H}_{2}^r&=(-3.4\pm 1.5)\times10^{-3}\;.
\label{LEC}
\end{align}
We will only use the central values of the three-flavor LECs for generating our plots since including the uncertainrties gives rise to a complex $\eta$-mass which is unphysical~\cite{usagain}.
We get the following bare parameters
\begin{align}
m_{\pi,0}^{\rm cen}&=131.28\;{\rm MeV}\;,\;
  &m_{K,0}^{\rm cen}=520.65\;{\rm MeV}\;,\\
f_{\pi,0}^{\rm cen}&=75.16\;{\rm MeV}\;,\\
m_{\pi,0}^{\rm low}&=128.14\;{\rm MeV}\;,\;
  &m_{K,0}^{\rm low}=512.72\;{\rm MeV}\;,\\
f_{\pi,0}^{\rm low}&=75.68\;{\rm MeV}\;,\\
m_{\pi,0}^{\rm high}&=134.43\;{\rm MeV}\;,\;
  &m_{K,0}^{\rm high}=528.76\;{\rm MeV}\;,\\
f_{\pi,0}^{\rm high}&=77.62\;{\rm MeV}\;.
\end{align}
Similarly, the experimentally determined two-flavor LECs used to generate the two-flavor condensates are
\begin{align}
\label{LEC2old}
\bar{l}_{1}&=-0.4\pm0.6\;,\\
\bar{l}_{2}&=4.3\pm0.1\;,\\
\bar{l}_{3}&=2.9\pm2.4\;,\\
\bar{l}_{4}&=4.4\pm0.2\;,\\
\bar{h}_1&=-1.5\pm0.2\;.
\end{align}
These are proportional to the running LECs evaluated at the bare pion mass as follows from their definitions in Eqs. (\ref{2fLECl}) and (\ref{2fLEC})

\subsection{Deviation of condensates at $j=0$}

\begin{figure}[htb]
\includegraphics[width=0.45\textwidth]{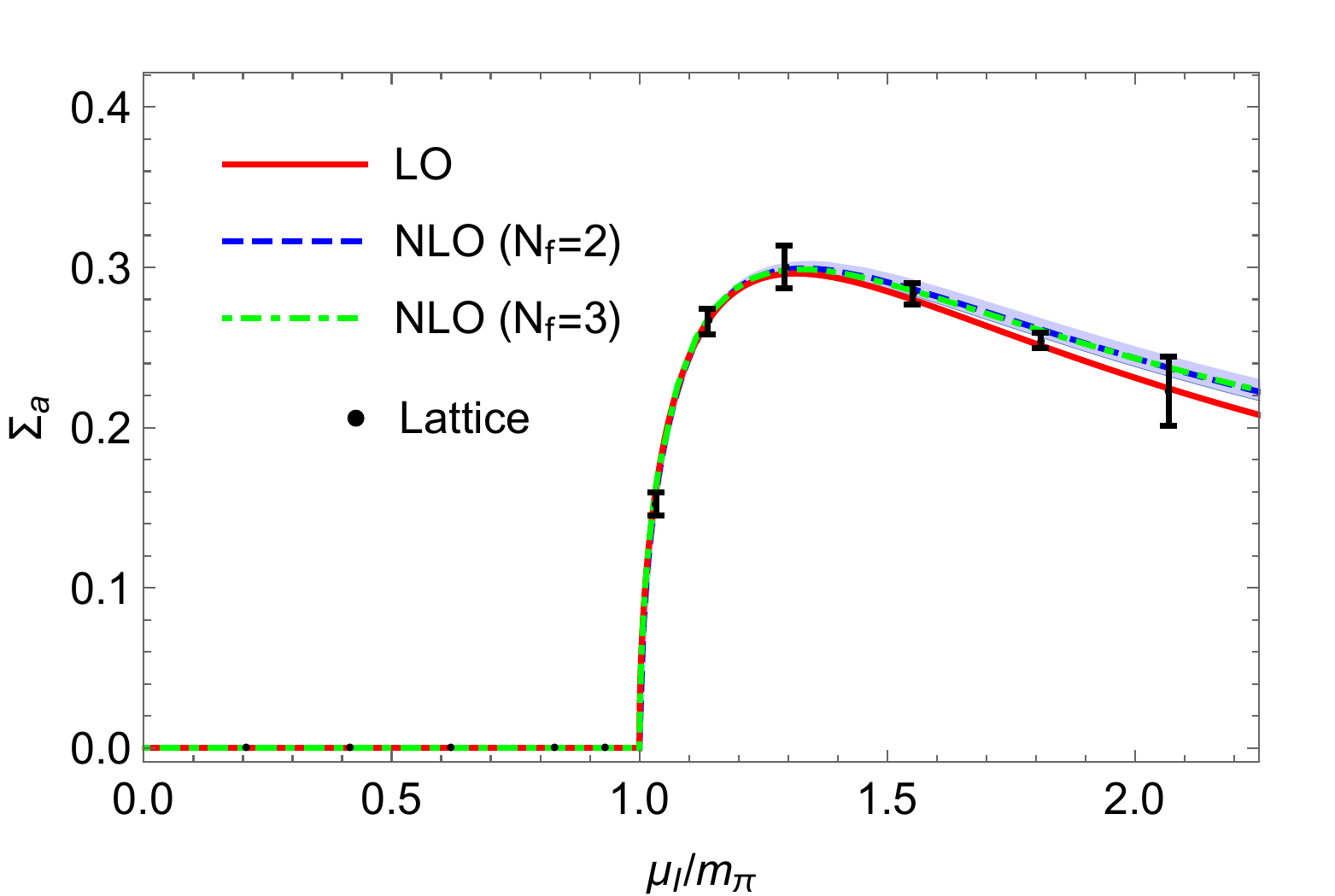}
  \caption{Axial condensate deviation, $\Sigma_{a}$, as a function of the isospin chemical potential at tree level and NLO for $j=0$. See main text for details.}
  \label{fig:ac}
\end{figure}

In Fig.~\ref{fig:ac}, we plot the axial condensate deviation, which is the negative of the axial condensate, at tree level and NLO. We find that both the tree-level and the NLO axial condensates are in excellent agreement with lattice QCD. The difference between the tree-level, NLO and lattice is negligible up to $\mu_{I}\approx 1.2m_{\pi}$ with the differences becoming more significant with increasing isospin chemical potentials. The difference between the two-flavor and the three-flavor result is tiny. 

\begin{figure}[htb]
\includegraphics[width=0.45\textwidth]{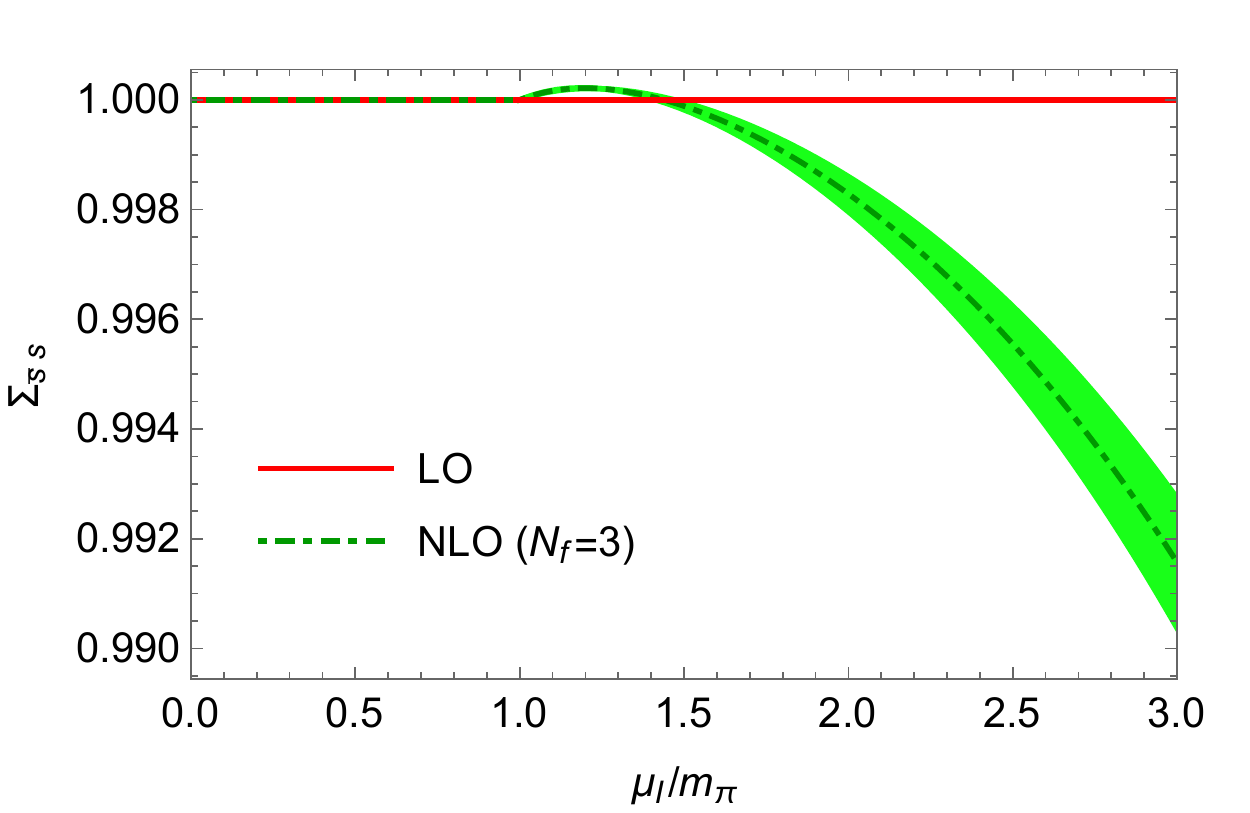}
  \caption{Deviation of the strange-quark condensate (normalized to 1) from the normal vacuum value, $\Sigma_{\bar{s}s}$, in three-flavor $\chi$PT for $j=0$. See main text for details.}
  \label{fig:sc}
\end{figure}

In Fig.~\ref{fig:sc}, we plot the strange-quark condensate deviation at both tree level (red) and next-to-leading order (green). At tree level, the pion condensate does not expel the strange-quark condensate. However, at NLO, the deviation of the strange-quark condensate increases above one up to approximately $\mu_{I}=1.4m_{\pi}$ and then decreases monotonically relative to its vacuum value. Compared to the light-quark condensate, the decrease is significantly smaller. Note that the normalizations in the chiral and quark condensate deviations are different by factors of $f_{\pi}^{2}$ and $f_{K}^{2}$ respectively, which are insufficient to explain the difference in the deviations of the respective condensates.
It would be of interest to calculate the strange-quark condensate on the lattice
to see if it displays the non-monotonic behavior found here.

\subsection{Deviation of condensates at $j\neq 0$}
\begin{figure}[htb]
\centering
\includegraphics[width=0.45\textwidth]{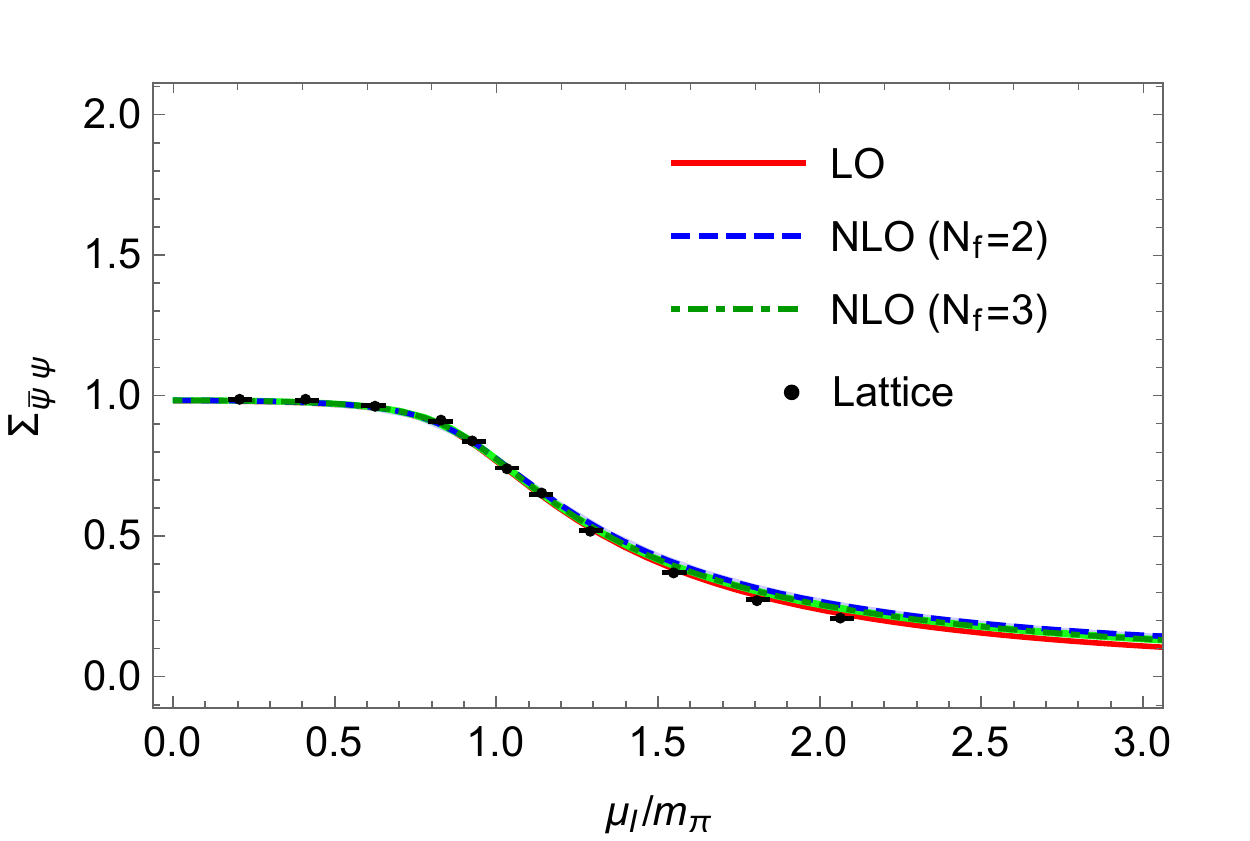}\,\,\,\,\,
\includegraphics[width=0.45\textwidth]{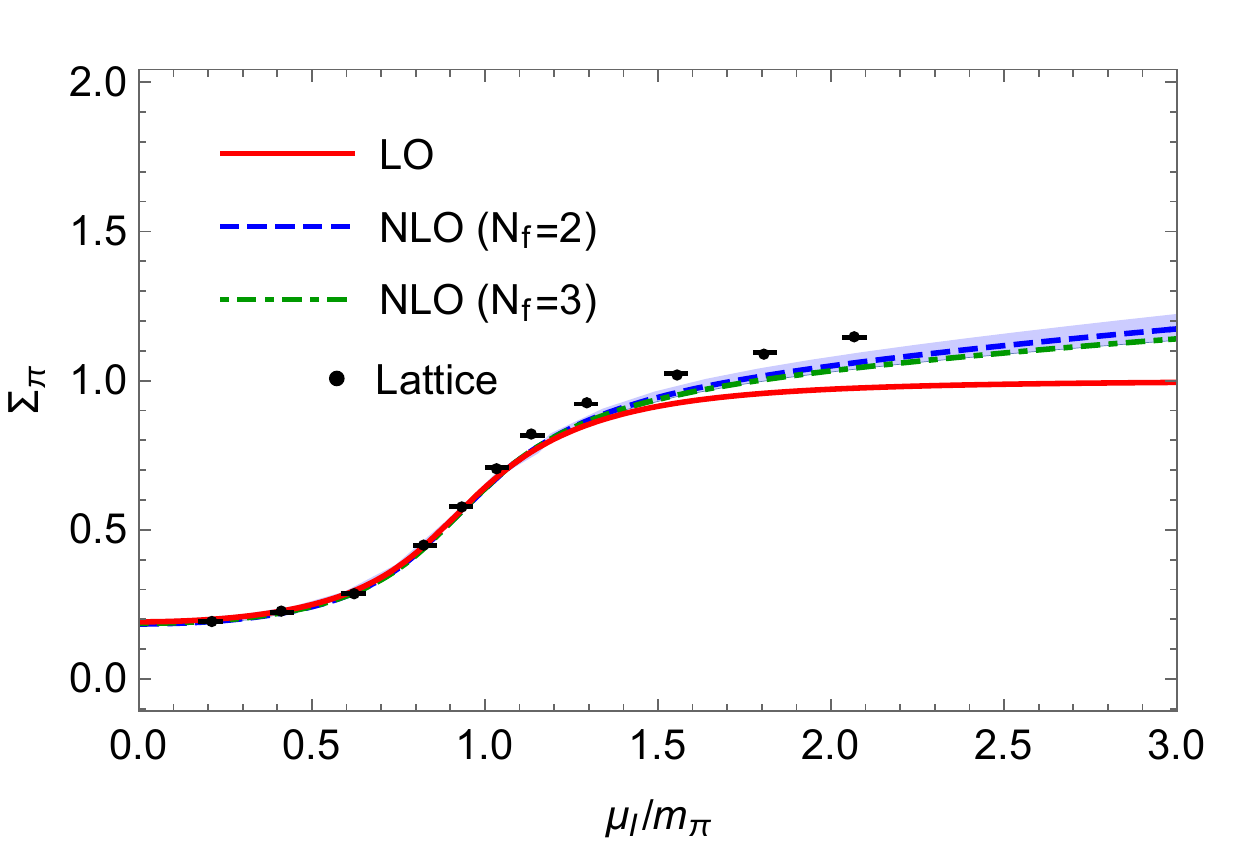}\,\,\,\,\,
\caption{Upper panel shows the deviation of the 
light quark  condensate (normalized to $1$) from the vacuum value, $\Sigma_{\bar{\psi}\psi}$ for $j=0.00517054m_{\pi}$. Lower panel shows the
deviation of the pion condensate from the vacuum value, $\Sigma_{\pi}$ for  $j=0.00517054m_{\pi}$. See main text 
in~\cite{pionstar,gergy3} for details.}
\label{fig:ccpcj1}       
\end{figure}

\begin{figure}[htb]
\centering
\includegraphics[width=0.45\textwidth]{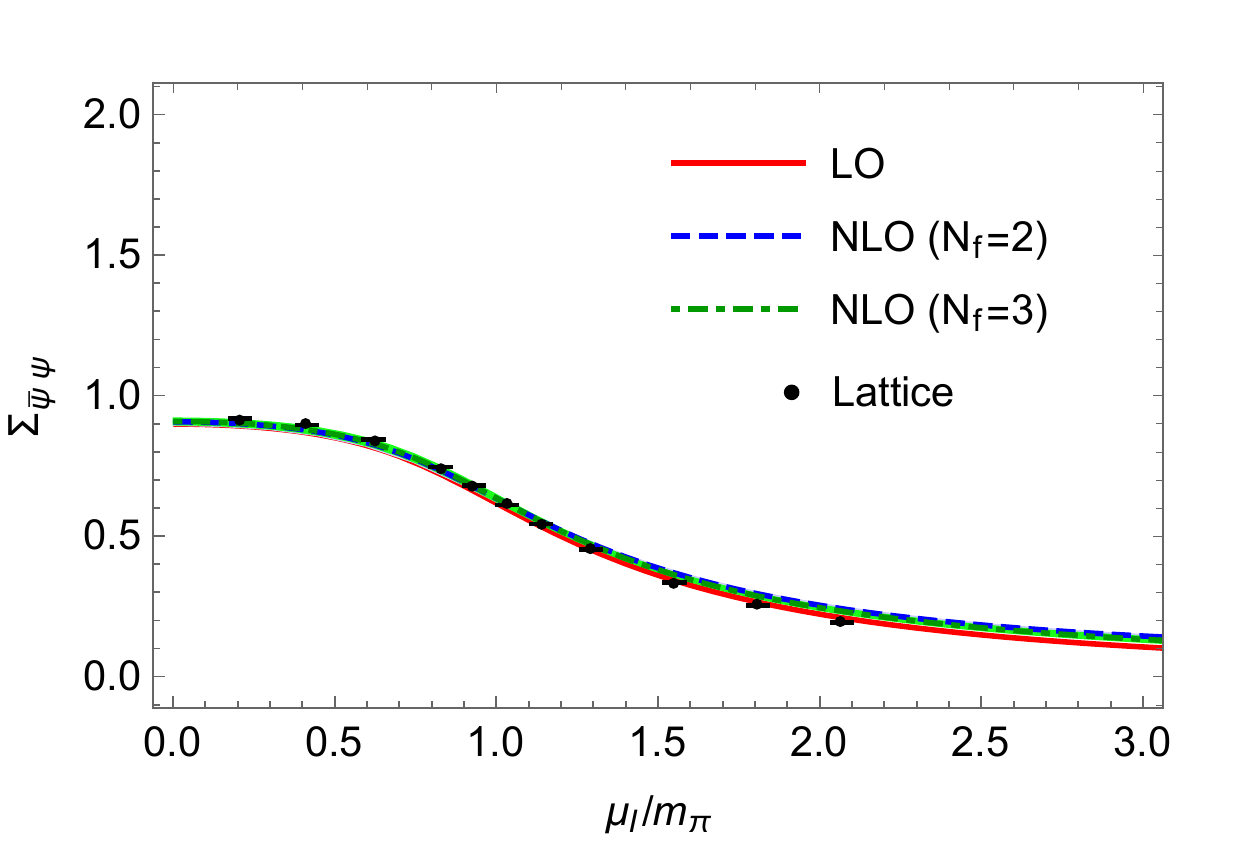}\,\,\,\,\,
\includegraphics[width=0.45\textwidth]{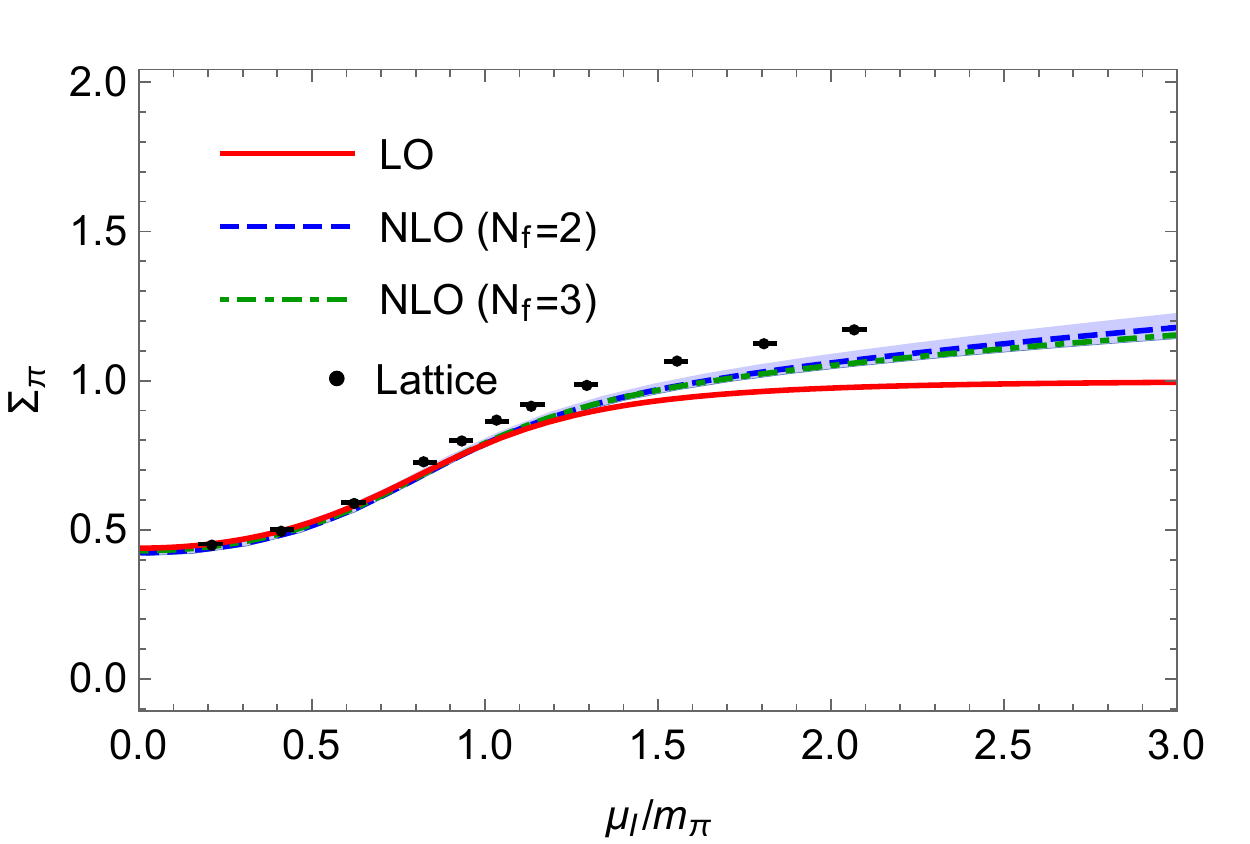}\,\,\,\,\,
\caption{Upper panel shows the deviation of the light quark  condensate (normalized to $1$) from the vacuum value, $\Sigma_{\bar{\psi}\psi}$, for $j=0.0129263m_{\pi}$. Lower panel shows the deviation of the pion condensate from the vacuum value, $\Sigma_{\pi}$, for $j=0.0129263 m_{\pi}$. See main text in~\cite{pionstar,gergy3} for details.}
\label{fig:ccpcj2}       
\end{figure}

In this subsection, we compare $\chi$PT light quark and pion condensates at finite $j$ with available QCD lattice data~\cite{pionstar,private,latiso}. In Fig.~\ref{fig:ccpcj1}, we show the deviation of the chiral and pion condensates
as defined in Eq.~(\ref{deviation}) for $j=0.00517054 m_{\pi}$, which is the smallest value of the source for which lattice data is available. In Fig.~\ref{fig:ccpcj2}, we show the deviation of the chiral and pion condensates for $j=0.0129263 m_{\pi}$. We note that there is no chiral and pion condensate data available for $j=0$ since they are ``cumbersome" to generate~\cite{gergy3}. 

For a fair comparison of the finite $j$ lattice data, it is important to know the quark masses in the continuum. Since quark masses are not physical observables their values depend on the method of renormalization. For the lattice calculation, a continuum extrapolation was not performed. Consequently, we use the lattice continuum quark masses of Ref.~\cite{BMW} for our comparison (and include a $5\%$ uncertainty) with the expectation that the difference with the lattice calculation of Ref.~\cite{pionstar,latiso} is small.

The upper panel of Fig.~\ref{fig:ccpcj1} shows the light quark  condensate deviation at $j=0.00517054m_{\pi}$ from $\chi$PT and lattice QCD as a function of $\mu_I/m_{\pi}$. 
Firstly, we observe that the NLO correction to the LO results (red solid line) 
is very small for both $N_f=2$ (blue dashed line) and $N_f=3$ (green dashed line).
All three curves are in excellent agreement with the lattice results (black points),
the tree-level results being in slightly better agreement.
In the lower panel of Fig.~\ref{fig:ccpcj1}, we plot the pion condensate deviation for the same value of
the pionic source. The pion condensate is therefore nonzero for all values of
$\mu_I$ because the nonzero pseudo-scalar source explicitly breaks isospin symmetry.
The tree-level and NLO pion condensate agree with each other and the lattice results up to $\mu_{I}\approx 1.2m_{\pi}$. Beyond that $\chi$PT underestimates the pion condensate with two-flavor $\chi$PT in better agreement with lattice QCD compared to three-flavor $\chi$PT. Notice that the LO results level off for large
values of $\mu_I$ independent of the source $j$, in disagreement
with both lattice data and the NLO results. Thus the NLO result is a significant
improvement over the tree-level result and we can no longer interpret $\alpha$ as the angle specifying how the chiral condensate is rotated into the pion condensate. A similar violation is seen in the NJL model~\cite{NJLisospin}.

In Fig.~\ref{fig:ccpcj2}, we plot the light quark and pion condensate deviations at $j=0.0129263 m_{\pi}$ from $\chi$PT and lattice QCD. The qualitative behavior is similar to that for $j=0.00517054m_{\pi}$ and the same remarks apply,
in particular the improved agreement of the chiral condensate
with lattice data.




\section*{Acknowledgements}
The authors would like to thank B. Brandt, G. Endr\H{o}di and S. Schmalzbauer
for providing their lattice condensate data~\cite{pionstar}. 
P.A. would also like to acknowledge helpful discussions with B. Brandt.
\appendix

\section{Integrals}
We use dimensional regularization to regulate ultraviolet divergences.
With dimensional  regularization,  the momentum  integrals are
generalized  to $d=3-2\epsilon$ dimensions.
We use the following notation
\bqa
\int_{p}&=&\left (\frac{e^{\gamma_{E}}\Lambda^{2}}{4\pi} \right )^{\epsilon}
\int
\frac{d^{d}p}{(2\pi)^{d}}\;,
\label{int}
\eqa
where $\Lambda$ is the renormalization scale in the modified minimal
subtraction ($\overline{\rm MS}$) scheme.
The integral below is defined as in Eq.~(\ref{int}), but with $d=4-2\epsilon$
and the subscript $p$ being replaced by $P$
\bqa
\int_P&=&\int{dp_0\over2\pi}\int_p\;.
\eqa
The integral we need in order to regularize the one-loop effective potential is
\begin{align}\nonumber
&\int_P\log[P^2+m^2]
=\int_p\sqrt{p^2+m^2}
\\ 
&=-{m^4\over2(4\pi)^2}
\left({\Lambda^2\over m^2}\right)^{\epsilon}
\left[{1\over\epsilon}+{3\over2}+{\cal O}(\epsilon)\right]\;.
\label{sumint0}
\end{align}


\begin{thebibliography}{}
\bibitem{raja}
K. Rajagopal and F. Wilczek, 
At the frontier of particle physics, Vol. 3
(World Scientific, Singapore, p 2061) (2001).

\bibitem{alford}
M. G. Alford, A. Schmitt, K. Rajagopal, and T. Sch\"afer,
Rev. Mod. Phys. {\bf 80}, 1455 (2008).

\bibitem{fukurev}
K. Fukushima and  T. Hatsuda,
Rept. Prog. Phys. {\bf 74}, 014001 (2011).


\bibitem{Nambu} Y. Nambu and G. Jona-Lasinio, Phys. Rev.
  \textbf{122}, 345 (1961).

\bibitem{BCS} J. Bardeen, L.N. Cooper and J.R. Schrieffer, Phys. Rev.
  \textbf{106}, 162 (1957).
  

\bibitem{Scherer} S. Scherer, Adv. Nucl. Phys. \textbf{27}, 277 (2003).

\bibitem{mannarev}  
M. Mannarelli,
Particles {\bf 2}, 411 (2019). 

\bibitem{son} D. T. Son and M. A. Stephanov, Phys. Rev. Lett. \textbf{86}, 592
(2001).  
  

\bibitem{isomag} P. Adhikari, T. D. Cohen, J. Sakowitz,
Phys. Rev. \textbf{C} 91, 045202 (2015); P. Adhikari, Phys. Lett. B 790, 211 (2019).  



\bibitem{2fcondensate}P. Adhikari, and J.O. Andersen, 
    Eur. Phys. J. C {\bf 80}, 11 (2020).

\bibitem{Cohen} T. D. Cohen, R.J. Furnstahl and D.K. Griegel, Phys. Rev.
  \textbf{C} 45, 1881 (1992).  
%


\bibitem{pionstar}
B. B. Brandt, G. Endr\H{o}di, E. S. Fraga, M. Hippert, J. Schaffner-Bielich and S. Schmalzbauer, Phys. Rev. D \textbf{98}, 094510 (2018).

\bibitem{axialfirst} M. A. Metlitsky, and A. R. Zhitnitsky, Nucl. Phys. B \textbf{731}, 309  (2005).


\bibitem{NJLaxial} T. Brauner, and X. G. Huang, Phys. Rev. D \textbf{94}, 094003 (2016).



\bibitem{wein}
  S. Weinberg, 
  Physica A {\bf 96}, 327 (1979).
  

  
\bibitem{gasser1}
  J. Gasser and H. Leutwyler,
Ann. Phys. {\bf 158}, (142) (1984).

\bibitem{gasser2}
J. Gasser and H. Leutwyler, Nucl. Phys. B {\bf 250}, 465 (1985).

\bibitem{bein}
  J. Bijnens, G. Colangelo and G. Ecker,
  Ann. Phys. {\bf 280}, 100 (2000).
 

  
\bibitem{kogut3}	
J. B. Kogut and D. Toublan 
Phys. Rev. D {\bf 64}, 034007 (2001).


\bibitem{us}P. Adhikari and J. O. Andersen,
Phys. Lett. B {\bf 804}, 135352 (2020),
JHEP {\bf 06}, 170 (2020).

  
%
\bibitem{LoewefiniteT} M. Loewe and C. Villavicencio, Phys. Rev. D \textbf{71}, 094001 (2005).

%

\bibitem{kim}
K. Splittorff, D. T. Son, M. A. Stephanov,
Phys. Rev. D {\bf 64}, 016003 (2001).

\bibitem{usagain}
  P. Adhikari, J. O. Andersen,  and P. Kneshcke,
  Eur. Phys. J. C {\bf 79}, 874 (2019).

\bibitem{gergy3} 	
B. B. Brandt, G. Endr\H{o}di, and S. Schmalzbauer,
Phys. Rev. D {\bf 97}, 054514 (2018).


\bibitem{Hr2ref1}
M. Jamin, Phys. Lett. B \textbf{538}, 71 (2002).

\bibitem{Hr2ref2}
J. Bordes, C.A. Dominguez, P. Moodley, J. Pe$\tilde{\textrm{n}}$arrocha and K. Schilcher, JHEP \textbf{10}, 102 (2012).


\bibitem{private}G. Endr\H{o}di, private communication.

\bibitem{BMW}
BMW Collaboration, S. Durr, Z. Fodor, C. Hoelbling, S. D. Katz, S. Krieg, T. Kurth, L. Lellouch, T. Lippert, K.K. Szabo and G. Vulvert, Phys. Lett. B, \textbf{707}, 265 (2011).


\bibitem{bijnensreview}
J. Bijnens and G. Ecker,
Ann. Rev. Nucl. Part. Sci. {\bf 64}, 149 (2014).



  
  
  


  



  


\bibitem{latiso} B. B. Brandt, G. Endr\'{o}di, and S. Schmalzbauer, EPJ Web
  Conf. \textbf{175}, 07020 (2018); B. B. Brandt, G. Endr\'{o}di, and
  S. Schmalzbauer, Phys. Rev. D \textbf{97}, 054514 (2018).
  
\bibitem{NJLisospin} T. Xia, L. He, and P. Zhuang, Phys. Rev. D 88, 056013
  (2013); S. S. Avancini, A. Bandyopadhyay, D. C. Duarte, R. L. S. Farias, Phys.
  Rev. \textbf{D} 100, 116002 (2019).






















%
%
%
%
%



  
\end{thebibliography}
\end{document}